\documentclass[prd,onecolumn,notitlepage,showpacs,preprintnumbers,amsmath,amssymb,nofootinbib,APS,11pt,superscriptaddress]{revtex4-1}

\usepackage{dcolumn}
\usepackage{bm}
\usepackage[applemac]{inputenc}
\usepackage[spanish,english]{babel}
\usepackage{amsmath,amssymb,amsfonts,latexsym,cancel}
\usepackage{graphicx}
\usepackage{color}
\usepackage{soul}
\usepackage{ulem}
\usepackage{hyperref}
\newcommand{\be}{\begin{equation}}
\newcommand{\ee}{\end{equation}}
\newcommand{\bea}{\begin{eqnarray}}
\newcommand{\eea}{\end{eqnarray}}

\begin{document}
\title{{\bf Renormalized stress-energy tensor for spin-1/2 fields in expanding universes}}

\author{ Adrian del Rio}\email{adrian.rio@uv.es}
\author{Jose Navarro-Salas}\email{jnavarro@ific.uv.es}
\affiliation{Departamento de Fisica Teorica and IFIC, Centro Mixto Universidad de Valencia-CSIC. Facultad de Fisica, Universidad de Valencia, Burjassot-46100, Valencia, Spain.}
\author{Francisco Torrenti}\email{f.torrenti@csic.es}
\affiliation{Instituto de Fisica Teorica UAM/CSIC, Universidad Autonoma de Madrid, Cantoblanco, 28049 Madrid, Spain.}

\begin{abstract}

We provide an explicit expression for the renormalized expectation value of the stress-energy tensor of a spin-$1/2$ field in a spatially flat Fridmann-Lemaitre-Robertson-Walker universe. Its computation is based on the extension of the adiabatic regularization method to fermion fields introduced recently in the literature. The tensor is given in terms of UV-finite integrals in momentum space, which involve the mode functions that define the quantum state.  
As illustrative examples of the method efficiency, we see how to compute the renormalized energy density and pressure in two interesting cosmological scenarios: a de Sitter spacetime and a radiation-dominated universe. In the second case, we explicitly show that the late-time renormalized stress-energy tensor behaves as that of classical cold matter.   
We also check that, if we obtain the adiabatic expansion of the scalar field mode functions with a similar procedure to the one used for fermions, we recover the well-known WKB-type expansion.\\

{\it Key words:} quantum field  theory in curved spacetime, renormalization, cosmology.
\end{abstract}

\pacs{ 04.62.+, 11.10.Gh, 98.80.k, 98.80.Cq }

\date{August, 26 2014}

\maketitle

\section{Introduction}\label{Introduction}

One of the most important consequences of combining quantum theory with general relativity is the phenomenon of gravitational particle creation, as first discovered in \cite{parker66, parker68} (see also the reviews \cite{parker-toms,birrell-davies}). The generation and amplification of quantum field  fluctuations  is inevitable during the expansion of the universe, and hence the creation of quanta. Only conformally invariant fields in conformally invariant backgrounds are preserved from producing particles. Therefore, a Friedmann-Lemaitre-Robertson-Walker (FLRW) universe necessarily produces scalar particles (minimally coupled), massive fermions, tensorial perturbations (gravitational waves) and scalar perturbations. In order to create a significant amount of these particles, waves, and perturbations, we need rapid expansions, like those expected to happen in the very early universe \cite{inflation}. This is essentially the mechanism driving the generation of primordial inho
 mogeneities observed in the large-scale structure of the universe and in the cosmic microwave background \cite{inflation2}. Moreover, the recent results of the BICEP2 experiment \cite{bicep2}, if confirmed, can offer the first  evidence of gravitational wave creation by the early expanding universe.

The gravitationally produced particles or perturbations contribute to the energy density and pressure with new ultraviolet  (UV) divergences, not present in the quantization of free fields in Minkowski spacetime. Therefore, one needs to use a self-consistent regularization and renormalization scheme in curved spacetime to subtract these divergences properly. 

One of the most useful schemes in cosmological scenarios is the adiabatic regularization method. It was originally designed to overcome the UV divergences of the particle number operator \cite{parker66}. (For a very nice historical account see \cite{Parker12}.) Later, it was generalized and systematized to renormalize the vacuum expectation value of the stress-energy tensor \cite{parker-fulling, Bunch80, Anderson-Parker}.  In the case of scalar fields, this method is based on a WKB-type expansion of the field modes, which allows one to identify the divergent terms of the tensor unequivocally and to subtract them from the bare expressions. More specifically, each term in the momentum mode sum is compared with what would it be if the expansion of the universe was slowed down. An adiabatic expansion of these terms is carried out in terms of a ``slowness parameter'', and the first three terms (adiabatic orders $0$, $2$ and $4$) are subtracted. The final quantity is hence renormalized
 . We remark that the adiabatic subtraction provides as a final result finite integrals in momentum space, which means that the subtraction is also playing the role of a regularization procedure. This is the reason for naming ``adiabatic regularization'' to the whole process of renormalization. Moreover, as in the end one only needs to evaluate these finite integrals, the method is computationally very powerful. On the other hand, the subtracted terms can also be interpreted in terms of renormalization of coupling constants in the gravitational action functional \cite{Bunch80}. The renormalization scheme is therefore covariant, as a consequence of the fact that adiabatic invariance is actually a covariant concept. The adiabatic regularization scheme has also been used to scrutinize the power spectrum in inflationary cosmology  \cite{inflation-r} and the inflationary completion of quantum gravity \cite{inflation-qg}.

One of the main issues with the renormalization program in curved spacetime is that these methods have been mainly developed for free scalar bosons, and less work has been done for other fields. In particular, an adiabatic regularization method for spin-1/2 fields in an expanding universe was missing until very recently \cite{landete}. One of the main features of the extended method is that the UV-divergent terms of the different physical quantities are not identified through a WKB-type expansion of the field modes, but with a different expansion which was introduced in \cite{landete}. As a nontrivial test of the new method, we worked out the conformal anomaly, recovering exactly the results obtained by other renormalization prescriptions.  

In this paper, we apply the adiabatic regularization method to obtain a general and explicit expression for the renormalized stress-energy tensor of a spin-1/2 field in a FLRW universe. This result, given in Eqs. (\ref{18}), (\ref{t00ren}) and (\ref{tiiren}), is written in terms of UV-convergent momentum integrals involving the field modes. This is a necessary and unavoidable step to prepare the method to be used for numerical computations in cosmology.  As illustrative examples, we study the renormalized stress-energy tensor in de Sitter space and in a radiation-dominated universe. In both examples, we need to specify appropriate initial conditions in order to ensure the renormalizability properties of the tensor. We also prove here that the same procedure used in \cite{landete} to obtain the adiabatic expansion of the fermionic field modes leads to the well-known WKB-type expansion when the algorithm is applied for scalar modes. This confirms definitely the appropr
 iateness of
  the fermionic adiabatic regularization method.

The content of the present work is organized as follows: In Sec. II, we introduce the basic equations in a FLRW metric and the adiabatic regularization method for spin-1/2 fields. In Sec. III, we prove that the adiabatic expansion for fermion fields recovers the WKB-type expansion when applied to scalar fields. In Sec. IV, we work out explicitly the expressions of the vacuum expectation values of the different components of the fermion stress-energy tensor. We renormalize these components and prove that the stress-energy tensor in conserved. In Sec. V, we apply the method to two specific examples: de Sitter spacetime and radiation-dominated universe. In the second case, we confirm that the late-time behavior of the created particles recovers the state equation of cold matter. Finally, Sec. VI contains our main conclusions. We use units $c = \hbar = 1$, and the conventions in \cite{parker-toms, birrell-davies}. In particular, the metric signature is $(+,-,-,-)$.

\section{Quantized spin-$1/2$ fields and the adiabatic expansion}

 A spin-1/2 field $\psi$ of mass $m$ in curved spacetime is described by the Dirac equation
\be (i\underline{\gamma}^{\mu}\nabla_{\mu}-m)\psi=0 \ , \label{1}\ee
where $\underline{\gamma}^{\mu}(x)$ are the spacetime-dependent Dirac matrices satisfying the anticommutation relations $\{\underline{\gamma}^{\mu},\underline{\gamma}^{\nu}\}=2g^{\mu\nu}$, and $\nabla_{\mu}\equiv \partial_{\mu}-\Gamma_{\mu}$ is the covariant derivative associated to the spin connection $\Gamma_{\mu}$.

In a spatially flat FLRW universe,  $ds^2=dt^2-a^2(t)d\vec{x}^2$, the matrices $\underline{\gamma}^{\mu}(t)$ are related with the constant Minkowskian matrices $\gamma^{\alpha}$ (which satisfy the relation $\{\gamma^{\alpha},\gamma^{\beta}\}=2\eta^{\alpha\beta}$) by $\underline{\gamma}^0(t)=\gamma^0$ and $\underline{\gamma}^i(t)=\gamma^i/a(t)$. On the other hand, the spin connections are in this metric $\Gamma_0 = 0$ and $\Gamma_i = (\dot{a}/2) \gamma_0 \gamma_i$. Therefore, $\underline{\gamma}^{\mu}\Gamma_{\mu}=-(3\dot a/2 a) \gamma_0$, and the differential equation (\ref{1}) can be written as
\bea
\left(i\gamma^0 \partial_0+\frac{3i}{2}\frac{\dot a}{a}\gamma^0+\frac{i}{a}\vec{\gamma}\cdot\vec{\nabla}-m\right)\psi=0 \ , \label{2}
\eea
where $\vec{\gamma} = (\gamma^1, \gamma^2, \gamma^3)$. Throughout this paper we shall work with the Dirac-Pauli representation for the Dirac matrices
\be 
\gamma^0 =
 \left( {\begin{array}{cc}
  I & 0  \\
  0 & -I  \\
  \end{array} } \right) \ , 
 \hspace{2cm} \vec\gamma = \left( {\begin{array}{cc}
  0 & \vec\sigma  \\
  -\vec\sigma & 0  \\
  \end{array} } \right) \ , 
  \ee
  where $\vec{\sigma} = (\sigma_1, \sigma_2, \sigma_3)$ are the usual Pauli matrices.
By extending the quantization procedure in Minkowski space  one can construct, for a given $\vec k$,  two independent spinor solutions  as 
 \bea
u_{\vec{k}\lambda}(x)=u_{\vec{k}\lambda}(t)e^{i \vec{k} \cdot \vec{x}}=\frac{e^{i \vec{k} \cdot \vec{x}}}{\sqrt{(2\pi)^3 a^3}}
\begin{pmatrix}
h_k^I(t) \xi_{\lambda}(\vec{k})  \\
h_k^{II}(t) \frac{\vec{\sigma}\cdot \vec{k}}{k} \xi_{\lambda}(\vec{k})   
\end{pmatrix},  \label{3}
\eea 
 where  $k\equiv |\vec k|$ and $\xi_{\lambda}$ is a constant and normalized two-component spinor $\xi_{\lambda}^{\dagger} \xi_{\lambda '}=\delta_{\lambda'\lambda}$.    
 In this decomposition, $h_k^I$ and $h_k^{II}$ are two particular time-dependent functions obeying from (\ref{2}) the following coupled differential equations,
\bea
h^{II}_k=\frac{i a}{k}(\partial_t+i m)h^{I}_k \ , \hspace{2cm} h^{I}_k=\frac{i a}{k}(\partial_t-i m)h^{II}_k   \ , \label{3b}
\eea
and the following two uncoupled second order differential equations,
\bea
\left(\partial_t^2+\frac{\dot a}{a}\partial_t+i m\frac{\dot a}{a}+m^2+\frac{{k}^2}{a^2} \right)h_k^{I} & = & 0  \ ,  \label{3c}\\
\left(\partial_t^2+\frac{\dot a}{a}\partial_t-i m\frac{\dot a}{a}+m^2+\frac{{k}^2}{a^2} \right)h_k^{II} & = & 0  \ . 	   \label{3d}
\eea
For our purposes, it is convenient to use helicity eigenstates  $\xi_{\lambda}(\vec k)$, which follow the property  $\frac{\vec{\sigma}\vec{k}}{2k}\xi_{\lambda}(\vec{k})=(\lambda/2) \xi_{\lambda}(\vec{k})$, where $\lambda/2= {\pm}1/2$ represent the eigenvalues for the helicity. Their explicit form is ($\vec{k} = (k_1,k_2,k_3)$)
\be \xi_{+1}(\vec k)= \frac{1}{\sqrt{2k(k+k_3)}}\begin{pmatrix}
k+k_3  \\
k_1+ik_2   
\end{pmatrix} \ , \hspace{2cm} \xi_{-1}(\vec k)= \frac{1}{\sqrt{2k(k+k_3)}}\begin{pmatrix}
-k_1+ik_2  \\
k + k_3  
\end{pmatrix} \ . \label{spinores}\ee

Given a particular solution $(h^I_k(t), h_k^{II}(t))$ to  Eq. (\ref{3b}), one can naturally construct a new solution to the same equations $(-h^{II*}_k(t), h^{I*}_k(t))$.  
In Minkowski space, this is equivalent to jumping from a positive frequency solution to a negative frequency one. Therefore, one can construct two more independent and orthogonal solutions as 
\bea
v_{\vec{k}\lambda}(x)=\frac{e^{-i \vec{k} \cdot \vec{x}}}{\sqrt{(2\pi)^3 a^3}}
\begin{pmatrix}
-h_k^{II*}(t) \xi_{-\lambda}(\vec{k})  \\
-h_k^{I*}(t)  \frac{\vec{\sigma}\cdot \vec{k}}{k}\xi_{-\lambda}(\vec{k})   
\end{pmatrix} \ .   \label{4d}
\eea 
Using this approach, the $v_{\vec{k}\lambda}(x)$ modes are obtained by the charge conjugation operation: $v_{\vec{k} \lambda}=u_{\vec{k} \lambda}^c\equiv C\left[\bar u_{\vec{k} \lambda}\right]^{T}=i \gamma^2u_{\vec{k} \lambda}^* $. Note  that $-i\sigma_2\xi_{\lambda}^*(\vec k)=\lambda \xi_{-\lambda}(\vec k)$.
It is easy to check that $u^{\dagger}_{\vec{k}\lambda}v_{-\vec{k}\lambda'}=0$ and $(u_{\vec{k}\lambda},v_{\vec{k}\,'\lambda'})=0$, where 
the Dirac scalar product is given by \bea
(\psi_1,\psi_2)=\int d^3x a^3 \psi_1^{\dagger} \psi_2 \ .    \label{3e}
\eea
The normalization condition for the four-spinors, $(u_{\vec{k}\lambda},u_{\vec{k}\,'\lambda'})=(v_{\vec{k}\lambda},v_{\vec{k}\,'\lambda'})=\delta_{\lambda\lambda'}\delta^{(3)}(\vec{k}-\vec{k}\,') $  leads to
\bea
|h_k^I(t)|^2+|h_k^{II}(t)|^2 & = & 1 \ . \label{4}
\eea
This condition guarantees the standard anticommutation relations for the creation and annihilation operators $B_{\vec{k} \lambda}$ and $ D_{\vec{k} \lambda}$, 
defined by the expansion of the Dirac field in terms of the spinors defined above
\bea
\psi(x)=\int d^3\vec{k} \sum_{\lambda}\left[B_{\vec{k} \lambda}u_{\vec{k} \lambda}(x)+D_{\vec{k} \lambda}^{\dagger}v_{\vec{k} \lambda}(x) \right] \ .  \label{4c}
\eea

 \subsection{Adiabatic expansion}
 
The adiabatic regularization method for spin-1/2 fields, introduced in \cite{landete}, is based on the following ansatz for the field modes:
\be h^{I}_{{k}}(t) \sim \sqrt{\frac{\omega + m}{2\omega}} e^{-i \int^{t'} \Omega (t') dt'} F(t) \ , \,\,\,\,\,\, \,\,\,\,\,\, 
h^{II}_{{k}}(t) \sim \sqrt{\frac{\omega - m}{2\omega}} e^{-i \int^{t'} \Omega (t') dt'} G(t)\ , \label{fermion-ansatz} \ee
where $\omega \equiv \sqrt{(k/a(t))^2 + m^2}$ is the frequency of the mode and the  time-dependent functions $\Omega (t)$, $F(t)$ and $G(t)$ are expanded adiabatically as
\bea \Omega (t) &=& \omega + \omega^{(1)} + \omega^{(2)} + \omega^{(3)} + \omega^{(4)} + \dots \ ,  \nonumber \\
F (t) &=& 1 + F^{(1)} + F^{(2)} + F^{(3)} + F^{(4)} + \dots \ ,  \nonumber \\
G (t) &=& 1 + G^{(1)} + G^{(2)} + G^{(3)} + G^{(4)} + \dots \ . \label{adifexp}\eea
Here, $\omega^{(n)}$, $F^{(n)}$ and $G^{(n)}$ are functions of adiabatic order $n$, which means that they contain $n$ derivatives of the scale factor (for example, $\dot a $ is of adiabatic order 1 and $\ddot a \dot a^2$ is of adiabatic order 4). In the expansions above, we impose $F^{(0)} = G^{(0)} \equiv 1$ and $\omega^{(0)} \equiv \omega$ to recover the Minkowskian solutions in the adiabatic regime. 

In order to obtain the different terms of (\ref{adifexp}), we substitute (\ref{fermion-ansatz}) into the Dirac equations (\ref{3b}) and the normalization condition (\ref{4}). We have then the following system of three equations:
\bea 
\Omega F + i \dot{F} + i \frac{F}{2} \frac{d \omega}{d t} \left[ \frac{1}{\omega + m} - \frac{1}{\omega} \right] - m F &=& (\omega - m ) G \ ,  \nonumber \\
\Omega G + i \dot{G} + i \frac{G}{2} \frac{d \omega}{d t} \left[ \frac{1}{\omega - m} - \frac{1}{\omega} \right] + m G &=& (\omega + m ) F \ ,  \nonumber \\
(\omega + m) F F^{*} + (\omega - m ) G G^{*} &=& 2 \omega\label{fermsys} \ . 
\eea
We can obtain expressions for $F^{(n)}$, $G^{(n)}$ and $\omega^{(n)}$ by substituting (\ref{adifexp}) into (\ref{fermsys}) and solving the system order by order. In this process, we need to treat independently the real and imaginary parts of $F^{(n)}$ and $G^{(n)}$. This is done in detail in \cite{landete}. The expressions obtained contain ambiguities, which do not appear in the final renormalized physical quantities $\langle \bar{\psi} \psi \rangle$ and $\langle T_{\mu \nu} \rangle$. For the sake of simplicity, one can impose order by order the additional simplifying condition $F^{(n)} (m) = G^{(n)} (-m)$, which removes the spurious ambiguities and is a natural choice due to the symmetries of the equations of motion (\ref{3b}) under the change of the mass sign. With this, one obtains for the first two orders
 \bea  \omega^{(1)}&=& 0 \label{ferm-omega1} \ , \\ 
  F^{(1)}&=& -i\frac{m\dot a}{4  \omega^2 a} \label{effe1}  \ ,  \eea
and
\bea  \omega^{(2)}&=& \frac{5 m^4 \dot{a}^2}{8 a^2 \omega^5} - \frac{3 m^2 \dot{a}^2}{8 a^2 \omega^3} - \frac{m^2 \ddot{a}}{4 a \omega^3}  \ , \\
  F^{(2)}&=& - \frac{5 m^4 \dot{a}^2}{16 a^2 \omega^6} + \frac{5 m^3 \dot{a}^2}{16 a^2 \omega^5} + \frac{3 m^2 \dot{a}^2}{32 a^2 \omega^4} - \frac{m \dot{a}^2}{8 a^2 \omega^3} + \frac{m^2 \ddot{a}}{8 a \omega^4} - \frac{m \ddot{a}}{8 a \omega^3} \label{effe2} \ . \eea 
The third and fourth order contributions are written in Appendix A for completeness.

The adiabatic renormalization method consists in expanding adiabatically the momentum integral of the quantity we want to renormalize using (\ref{fermion-ansatz}), and subtracting enough adiabatic terms in order to ensure its convergence in the UV regime. The renormalization of the two-point function at coincidence  requires subtraction up to second order, while the stress-energy tensor needs subtraction up to fourth order. We apply this procedure to the renormalization of the stress-energy tensor in Sec. IV. First, to gain physical intuition and for readers more familiarized with the WKB expansion for scalar modes, we see below that a similar technique to the one used here can be equally applied for scalar fields. We rediscover this way the standard WKB-type adiabatic expansion.

\section{Another view on the adiabatic expansion for scalar fields}
In this section, we provide a new  view on the adiabatic expansion for scalar modes. Mimicking the  procedure designed to deal with fermions, we will recover the well-known bosonic WKB adiabatic expansion  without assuming it as an {\it a priori} input. Since the scalar field is more easy to manage, it can serve  to illustrate the prescription used for fermions. A free scalar field of mass $m$ in curved spacetime is described by the wave equation
\be ( \nabla^{\mu} \nabla_{\mu} + m^2 + \xi R ) \phi = 0 \ ,  \ee
where $\xi$ is the coupling of the field to the scalar curvature $R$. In the case of the spatially flat FLRW metric, the equation takes the form
\be \frac{\partial^2 \phi}{\partial t^2} + 3 \frac{\dot{a}}{a} \frac{\partial \phi}{\partial t} + \frac{1}{a^2} \sum_i \frac{\partial^2 \phi}{\partial {x^{i}}^2} + (m^2 + \xi R) \phi = 0 \ , \label{phi-eq} \ee
with $R=6 (\ddot{a}/a) + 6(\dot{a}^2 / a^2)$. One now expands the field as
\be \phi (\vec{x}, t) = \int \frac{d^3 \vec{k}}{\sqrt{2 (2 \pi a (t))^{3} }} (A_{\vec{k}} e^{i \vec{k} \vec{x}} h_k (t) + A_{\vec{k}}^{\dagger} e^{-i \vec{k} \vec{x}} h_k^{*} (t) ) \ , \label{phiexp}\ee
where $h_k (t)$ are time-dependent functions and the commutation relations for the creation and destruction operators are $[A_{\vec{k}}, A_{\vec{k'}}^{\dagger}] = \delta^{(3)} (\vec{k} - \vec{k'})$,$\ $ $[A_{\vec{k}}^{\dagger}, A_{\vec{k'}}^{\dagger}] = 0$, and $[A_{\vec{k}}, A_{\vec{k'}}] = 0$. If we substitute (\ref{phiexp}) into (\ref{phi-eq}), we find that $h_k (t)$ obeys the differential equation
\be \frac{d^2 h_k}{d t^2} + (\omega_k^2 (t) + \sigma) h_k = 0 \ , \label{hqeq} \ee
with $\omega_k (t) \equiv \sqrt{(k/a(t))^2 + m^2}$ and $\sigma \equiv (6 \xi - 3/4) (\dot{a} / a)^2 + (6 \xi - 3/2) \ddot a / a$. On the other hand, the right normalization condition for scalar fields is
\be h_k  \dot{h}_k^{*} - h_k^* \dot{h}_k = 2 i \ . \label{Wronskian}\ee

As for our analysis for spin-$1/2$ fields, we assume the following generic ansatz for the adiabatic expansion of the mode functions $h_k (t)$,
\be h_k (t) = H(t) e^{-i \int^t \Omega (t') dt'} \ , \label{ansatzWKB2} \ee
where here  $H(t)$  and $\Omega (t)$ are real functions. This simplifying assumption is somewhat equivalent to the natural symmetry relation used for spin-$1/2$ fields $F^{(n)} (m) = G^{(n)} (-m)$. We can expand them adiabatically as 
\be H(t) = \frac{1}{\sqrt{\omega_k}} + H^{(1)} (t) + H^{(2)} (t) + H^{(3)} (t) + H^{(4)} (t) + \dots\ ,  \label{hexp}\ee
and
\be \Omega (t) = \omega_k + \omega^{(1)} (t) + \omega^{(2)} (t) + \omega^{(3)} (t) + \omega^{(4)} (t) + \dots \label{omegaexp} \ . \ee
As for fermions, we have ensured that the zeroth order term of the expansion recovers the Minkowskian solutions: $H^{(0)} (t) \equiv \omega_k^{-1/2} (t)$ and $\omega^{(0)} (t) \equiv \omega_k (t) \equiv \omega (t) $. By substituting the ansatz (\ref{ansatzWKB2}) into the field equation (\ref{hqeq}) and the Wronskian (\ref{Wronskian}), we obtain the following system of two equations:
\bea 
\ddot{H} - H \Omega^2 - 2 i \Omega \dot{H} - i H \dot{\Omega} + (\omega^2 + \sigma) H &=& 0 \ , \nonumber \\
\Omega H^2 &=& 1 \ . \label{bos-sys}
\eea
Substituting (\ref{hexp}) and (\ref{omegaexp}) into (\ref{bos-sys}) and solving the system order by order, we find that the first term of both expansions is null, $\omega^{(1)} = H^{(1)} = 0$, and that the second order terms are
\be \omega^{(2)} (t) = \frac{5 m^4 \dot{a}^2}{8 a^2 \omega^5} - \frac{m^2 \dot{a}^2}{2 a^2 \omega^3} - \frac{\dot{a}^2}{2 a^2 \omega} + \frac{3 \xi \dot{a}^2}{a^2 \omega} - \frac{m^2 \ddot{a}}{4 a \omega^3} - \frac{\ddot{a}}{2 a \omega} + \frac{3 \xi \ddot{a}}{a \omega} \ , \ee
and
\be H^{(2)} (t) = - \frac{5 m^4 \dot{a}^2}{16 a^2 \omega^{13/2}} + \frac{m^2 \dot{a}^2}{4 a^2 \omega^{9/2}} + \frac{\dot{a}^2}{4 a^2 \omega^{5/2}} - \frac{3 \xi \dot{a}^2}{2 a^2 \omega^{5/2}} + \frac{m^2 \ddot{a}}{8 a \omega^{9/2}} + \frac{\ddot{a}}{4 a \omega^{5/2}} - \frac{3 \xi \ddot{a}}{2 a \omega^{5/2}} \ . \ee

This algorithm can be extended to all orders. One can immediately check that the expansions obtained this way are equivalent to the usual WKB-type expansions used for scalar fields \cite{parker-fulling}  
\be \label{ansatzWKB} h_k(t) = \frac{1}{\sqrt{W_k(t)}}e^{-i\int^t W_k(t')dt'} \ .  \ee
More specifically, one confirms that the expansion for $W_k(t)= \omega_k + \omega^{(1)}+ \omega^{(2)} + \dots$ is the same as the one for $\Omega_k(t)$ obtained above. One also sees that the $H^{(n)}$ are equal to
\be H^{(n)} = \left( \frac{1}{\sqrt{\omega_k + \omega^{(1)} + \omega^{(2)} + \dots} }\right)^{(n)} \ . \ee

One rediscovers this way the WKB expansion for scalar fields. The advantage of this strategy has been shown for spin-$1/2$ fields, as it seems that an efficient WKB-type adiabatic expansion for fermions in an expanding universe does not exist \cite{landete}.

\section{Renormalization of the  stress-energy tensor}

\subsection{Dirac stress-energy tensor components}

 The classical stress-energy tensor for a Dirac field in curved spacetime is given by
\bea
T_{\mu\nu}=\frac{i}{2}\left[\bar{\psi}\, \underline{\gamma}_{(\mu}\nabla_{\nu)}\psi-\bar{\psi}\  \overleftarrow{\nabla}_{(\nu}\, \underline{\gamma}_{\mu)}\psi \right]  \ ,  \label{10}
\eea
where $\psi$ is the Dirac field and $\underline{\gamma} (x)$ are the spacetime-dependent Dirac matrices. In the case of a FLRW universe, its homogeneity and spatial isotropy imply that we only have two independent components for this tensor: the energy density, related with the 00-component,  and the pressure, related with the ii-component. The 00-component can be written as
\be T_0^0 = \frac{i}{2} \left( \bar{\psi} {\gamma}^0 \frac{\partial \psi}{\partial t} - \frac{\partial \bar{\psi}}{\partial t} {\gamma}^0 \psi \right)\ ,  \label{ft00} \ee
while the ii-component is 
\be T_i^i  = \frac{i }{2a} \left( \bar{\psi} {\gamma}^i \frac{\partial \psi}{\partial x^i} - \frac{\partial \bar{\psi}}{\partial x^i} {\gamma}^i \psi \right) \ , \label{ftii} \ee
(not sum on $i$ implied). The former is directly computed  using $\Gamma_0=0$ and $\underline{\gamma}_0(x)=\gamma_0$. The latter is obtained  taking into account that $\Gamma_i=\frac{\dot a}{2}{\gamma}_0 {\gamma}_i$ and $\underline{\gamma}_i(x)=\gamma_i/a(t)$.

The next step is to compute the formal vacuum expectation values of the quantized stress-energy tensor. To this end, we will use the expansion of the Dirac field (\ref{4c}) in terms of the creation and annihilation operators. 
As a necessary previous result, we first compute the quantity $\left< \bar \psi \gamma^{\mu}\partial_{\nu}\psi\right>$. It is given by
\bea
  \left< \bar \psi \gamma^{\mu}\partial_{\nu}\psi\right>
  =  \int d^3\vec{k} \sum_{\lambda=\pm 1} \left(\bar v_{\vec{k}\lambda}\gamma^{\mu} \partial_{\nu}v_{\vec{k}\lambda} \right)  \ . \label{A3}
\eea 
With this result and with Eq. (\ref{4d}), we will compute the vacuum expectation value of (\ref{ft00}) and (\ref{ftii}) and obtain the corresponding expectation values for the energy density and pressure operators.

\subsection{Renormalized energy density}

Let us start with the energy density. If we take the expectation value of (\ref{ft00}) over the vacuum and use (\ref{A3}) and (\ref{4d}), we get after some algebra
\bea
\left< T_{00}\right>=\frac{1}{2\pi^2 a^3}\int_{0}^{\infty} dk k^2 \rho_k  \ , \label{12}
\eea
where
\be \rho_k (t) \equiv i \left( h_k^{I} \frac{\partial h_k^{I*}}{\partial t} + h_k^{II} \frac{\partial h_k^{II*}}{\partial t} - h_k^{I*} \frac{\partial h_k^{I}}{\partial t}  - h_k^{II*} \frac{\partial h_k^{II}}{\partial t} \right) \ . \label{13}\ee
Expression (\ref{12}) contains quartic, quadratic and logarithmic ultraviolet divergences, and consequently,  we must expand its integrand adiabatically and subtract from it enough terms of its expansion in order to have a finite quantity. By dimensionality, one expects to need subtraction up to fourth adiabatic order\footnote{It follows on dimensional grounds that if a quantity $Q$ has dimensions $M^d$ (where $M$ means mass), the $n$th adiabatic order term $Q^{(n)}$ in its expansion,
\be Q = Q^{(0)} + Q^{(1)} + Q^{(2)} + Q^{(3)} + Q^{(4)} + \dots  \ , \ee
 decays in the UV limit as $\mathcal{O} (k^{- \lambda})$ with $\lambda \geq \lambda_* \equiv n-d$. This can be confirmed by just looking at expressions (\ref{ferm-omega1})-(\ref{effe2}) and (\ref{omega3})-(\ref{effe4}) . For $Q =\rho_k$, we have dimension $d=1$, and we would require $\rho_k \sim k^{-4}$ in the UV limit in order to have (\ref{12}) finite. Therefore, as $n = \lambda_* + d = 4 + 1 = 5$, we need to subtract in (\ref{12}) all expansion terms from $\rho_k^{(0)}$ to $\rho_k^{(4)}$. This way, the first contribution to $\rho_k$ comes from the fifth order adiabatic terms.}.
To see this, we expand (\ref{13}) as
\be \rho_k = \rho_k^{(0)} + \rho_k^{(1)} + \rho_k^{(2)} + \rho_k^{(3)} + \rho_k^{(4)} + \dots   \ , \ee
where $\rho_k^{(n)}$ is of $n$th adiabatic order. In order to obtain the terms of the expansion, we substitute (\ref{fermion-ansatz}) and (\ref{adifexp}) into (\ref{13}) and obtain the contribution from the different adiabatic terms. We find that the zeroth adiabatic term corresponds to the usual Minkowskian divergence,
\be \rho_k^{(0)} = -2 \omega \ , \label{rho0} \ee
and that the first odd terms are null,
\be \rho_k^{(1)} = \rho_k^{(3)} = 0 \ . \ee
On the other hand, the second order term is given by
\be \rho_k^{(2)} = \frac{\omega + m}{\omega} \left( \mathfrak{Im} \dot{F}^{(1)} - |F^{(1)}|^2 \omega - 2 F^{(2)} \omega - \omega^{(2)} \right) + \frac{\omega - m}{\omega} \left( F \rightarrow G \right) \label{rhok2}\ ,  \ee
[$(F \rightarrow G)$ is the same expression as in the first parenthesis but changing $F$ by $G$] and the fourth-order term is
\bea \rho_k^{(4)} = && \frac{\omega + m}{\omega} \left( \mathfrak{Im} \dot{F}^{(3)} - \dot{F}^{(2)} \mathfrak{Im} F^{(1)} - (F^{(2)})^2 \omega - (F^{(3) *} F^{(1)} + F^{(1) *} F^{(3)} + 2 F^{(4)}) \omega \right. \nonumber \\
 &&  \left. +F^{(2)} (\mathfrak{Im} \dot{F}^{(1)} - 2 \omega^{(2)} ) - |F^{(1)}|^2 \omega^{(2)} - \omega^{(4)} \right) 
+ \frac{\omega - m}{\omega} \left( F \rightarrow G \right) \label{rhok4} \ . \eea
For completeness, the $n$th adiabatic order contribution to  (\ref{13}) is given by
\bea
\rho_k^{(n)}=-2\omega(t)^{(n)}+i\frac{\omega+m}{2\omega}\left[F \dot F^*-F^* \dot F   \right]^{(n)}+i\frac{\omega-m}{2\omega}\left[G \dot G^*-G^* \dot G   \right]^{(n)}.
\eea
In order to write (\ref{rhok2}) and (\ref{rhok4}) in terms of $\omega$ and the mass, we use (\ref{ferm-omega1})-(\ref{effe2}) and (\ref{omega3})-(\ref{effe4}). This gives
\be \rho_k^{(2)} = - \frac{m^4 \dot{a}^2}{4 \omega^5 a^2} + \frac{m^2 \dot{a}^2}{4 \omega^3 a^2} \label{rho2} \ , \ee 
and
\bea \rho_k^{(4)}  & = & \frac{105m^8 \dot a^4}{64\omega^{11} a^4} -\frac{91 m^6 \dot a^4}{32\omega^9 a^4}+\frac{81m^4 \dot a^4}{64\omega^7 a^4}-\frac{m^2 \dot a^4}{16\omega^5 a^4} -  \frac{7m^6 \dot a^2 \ddot a}{8\omega^9 a^3}+\frac{5m^4 \ddot a\dot a^2}{4\omega^7 a^3}\nonumber   \\
&-& \frac{3m^2 \dot a^2\ddot a}{8\omega^5 a^3}-\frac{m^4 \ddot a^2}{16\omega^7 a^2}  + \frac{m^2 \ddot a^2}{16\omega^5 a^2} +  \frac{m^4 \dot a\dddot a}{8\omega^7 a^2}-  \frac{m^2 \dot a \dddot a}{8\omega^5 a^2} \ . \nonumber\\
  \label{17}
\eea

The adiabatic renormalization subtraction terms are then defined as (we proceed in parallel to the case of scalar fields \cite{parker-fulling, parker-toms, birrell-davies})
\be \label{Adsubtractions}\langle T_{00} \rangle_{Ad} \equiv \frac{1}{2 \pi^2 a^3} \int_0^{\infty} dk k^2 (\rho_k^{(0)}+ \rho_k^{(2)} + \rho_k^{(4)} ) \ . \ee
Hence 
the renormalized 00-component of the stress-energy tensor is 
\be \langle T_{00} \rangle_{ren} \equiv   \langle T_{00} \rangle - \langle T_{00} \rangle_{Ad}=      \frac{1}{2 \pi^2 a^3} \int_0^{\infty} dk k^2 ( \rho_k - \rho_k^{(0)} - \rho_k^{(2)} - \rho_k^{(4)} ) \ . \label{ren-ten} \ee
This quantity is finite.

However, looking at expressions (\ref{rho2}) and (\ref{17}), one can observe that if we had subtracted only the terms up to second order, the tensor would already be convergent. In other words, the integral of the fourth-order adiabatic subtraction is, by itself, finite and independent of the mass of the field
\be   -\frac{1}{2 \pi^2 a^3} \int_0^{\infty} dk k^2  \rho_k^{(4)} = \frac{2}{2880\pi^2}\left[ -\frac{21}{2}\frac{\dot a^4}{a^4}+18\frac{\dot a}{a} \frac{\dddot a}{a}-9\frac{\ddot a^2}{a^2}+18\frac{\dot a^2}{a^2}\frac{\ddot a}{a}\right]  \ .   \label{27} \ee

Note that this also happens in the renormalization of a scalar field with conformal coupling $\xi=1/6$.  However, one must subtract up to the order necessary to remove the divergences for arbitrary values of $\xi$, and also for general metrics \cite{parker-toms}. For an arbitrary spacetime, the fourth adiabatic order contains real divergences, which disappear accidentally for FLRW metrics in the case of fermions or scalars with $\xi=1/6$ \cite{Christensen78}. Therefore, according to the general rule, we subtract up to fourth adiabatic order. Discarding the fourth adiabatic subtraction would lead to a vanishing trace anomaly [see Eq. (\ref{traceanomaly}) below].

\subsection{Renormalized pressure}

We can also derive the vacuum expectation value of the ii-component of the Dirac stress-energy tensor (\ref{ftii}) by direct computation using (\ref{A3}). Using  (\ref{4d}), one should arrive at the following expression
\bea
\left<\bar{\psi}\gamma^i\partial_i\psi \right>=\frac{i}{(2\pi)^3 a^3}\int d^3\vec{k} \sum_{\lambda= \pm 1} k_i \left[h_k^{I}h_k^{II*}+h_k^{I*}h_k^{II} \right]\lambda (\xi_{-\lambda}^{\dagger}\sigma^i \xi_{-\lambda})\, .
\eea
The property of isotropy of the FLRW spacetime allows us to perform the calculation for $i=3$ without loss of generality.  Therefore,
\bea
\left<\bar{\psi}\gamma^i\partial_i\psi \right>=\frac{i 2\pi}{(2\pi)^3 a^3}\int_{-1}^{1} d(\cos\theta) \cos \theta \sum_{\lambda= \pm 1} \lambda (\xi_{-\lambda}^{\dagger}\sigma^3 \xi_{-\lambda}) \int_0^{\infty} dk k^3  \left[h_k^{I}h_k^{II*}+h_k^{I*}h_k^{II} \right]  \label{pasointermedio} \ , 
\eea
where $\theta$ is the polar angle ($k_3=k \cos \theta$).
Using (\ref{spinores}), one finds
\bea
\sum_{\lambda= \pm 1} \lambda (\xi_{-\lambda}^{\dagger}\sigma^3 \xi_{-\lambda})=-2 \cos \theta\ , 
\eea
and plugging this into (\ref{pasointermedio}), the final result for the ii-component of the stress energy tensor reads 
\bea
\left< T_{ii}\right>=\frac{1}{2\pi^2 a}\int_{0}^{\infty} dk k^2 p_k  \ ,  \label{13c}
\eea 
with 
\bea
p_k\equiv-\frac{2k }{3a}[h_k^{I} h_k^{II*}+h_k^{I*} h_k^{II}]  \ . \label{13d}
\eea
Again, expression (\ref{13c}) contains several ultraviolet divergences, and consequently,  we must expand its integrand adiabatically and subtract from it enough terms of its expansion in order to have a finite quantity. Using the adiabatic expansion in (\ref{13d}), we get 
\bea
p_k^{(n)}=-\frac{\omega^2-m^2}{3 \omega}\left[F G^*+ F^* G\right]^{(n)} \ . \label{13e}
\eea
The corresponding renormalized $ii$-component is also defined as
\bea
\left< T_{ii}\right>_{ren}\equiv \left< T_{ii}\right> - \left< T_{ii}\right>_{Ad}= \frac{1}{2\pi^2 a}\int_{0}^{\infty} dk k^2 \left[p_k-p_k^{(0)} -p_k^{(2)}-p_k^{(4)} \right] \ ,  \label{17b}
\eea
with 
\be \left< T_{ii}\right>_{Ad}\equiv  \frac{1}{2\pi^2 a}\int_{0}^{\infty} dk k^2 \left[p_k^{(0)} +p_k^{(2)}+p_k^{(4)} \right]  \label{iiad} \ , \ee
and
$[p_k^{(1)}=p_k^{(3)}=0]$
\bea
p_k^{(0)} & = &-\frac{2}{3}\left[\omega-\frac{m^2}{\omega}\right],  \label{17c}\\
p_k^{(2)} & = & -\frac{m^2 \dot a^2}{12 \omega^3 a^2}-\frac{m^2 \ddot a}{6 \omega^3 a}+ \frac{m^4 \ddot a}{6 \omega^5 a}+\frac{m^4 \dot a^2}{2 \omega^5 a^2}-\frac{5 m^6 \dot a^2}{12 \omega^7 a^2},   \label{17d}\\
p_k^{(4)} & = & \frac{385   m^{10}\dot a^4}{64 \omega^{13}a^4} - \frac{791  m^8\dot a^4}{64 \omega^{11}a^4} + \frac{1477  m^6\dot a^4}{192 \omega^9a^4}- \frac{263  m^4\dot a^4}{192 \omega^7a^4}  +\frac{ m^2\dot a^4}{ 48 \omega^5 a^4} \nonumber\\
 &- & \frac{ 77  m^8 \dot a^2\ddot a}{16 \omega^{11}a^3} + \frac{77  m^6\dot a^2 \ddot a}{16 \omega^9a^3}  + \frac{175  m^6\dot a^2\ddot a}{ 48 \omega^9a^3}  - \frac{175  m^4\dot a^2\ddot a}{ 48 \omega^7a^3}- \frac{ m^4\dot a^2 \ddot a }{3 \omega^7a^3}+\frac{ m^2\dot a^2 \ddot a}{3 \omega^5 a^3}\nonumber\\
 & + & \frac{7  m^6\ddot a^2}{16 \omega^9a^2} -\frac{5  m^4\ddot a^2}{8 \omega^7 a^2} + \frac{3  m^2\ddot a^2}{16 \omega^5 a^2}  + \frac{7  m^6\dot a \dddot a}{12 \omega^9a^2} - \frac{5  m^4\dot a \dddot a}{6 \omega^7 a^2}  + \frac{ m^2\dot a \dddot a}{ 4 \omega^5 a^2}\nonumber \\
 &  & - \frac{ m^4\ddddot a}{24 \omega^7 a}+ \frac{ m^2\ddddot a}{24 \omega^5 a} \ .
\eea

We also note that the integral of the fourth-order subtraction terms is finite and mass independent: 
\be -\frac{1}{2\pi^2 a}\int_{0}^{\infty} dk k^2 p_k^{(4)} =  \frac{2a^2}{2880\pi^2}\left[ -\frac{7}{2}\frac{\dot a^4}{a^4}-12\frac{\dot a}{a} \frac{\dddot a}{a}-9\frac{\ddot a^2}{a^2}+14\frac{\dot a^2}{a^2}\frac{\ddot a}{a}-6\frac{\ddddot a}{a}\right] \\    \label{27b}\ . \ee

As a final comment, we stress that combining properly Eqs. (\ref{3b}), the following simple relationship between the pressure, the energy density  and the mode functions can be found:
\be 
\rho_k=3p_k-2m\left[|h_k^I|^2-|h_k^{II}|^2 \right]   \ , \label{17f}
\ee
where the second term in the right-hand side is basically $ \left<T_{\mu}^{\mu}\right>_k$, with
\bea
\left< T_{\mu}^{\mu}\right>=\frac{1}{2\pi^2 a^3}\int_{0}^{\infty} dk k^2 \left<T_{\mu}^{\mu}\right>_k  \ .
\eea
To see this, just remember that the trace is $\left< T_{\mu}^{\mu}\right>=\left< T_{00}\right>-\frac{3}{a^2}\left< T_{ii}\right>$, and then $\left<T_{\mu}^{\mu}\right>_k=\rho_k-3p_k \ .$

\subsection{Expression for the renormalized stress-energy tensor}

The  fourth-order adiabatic subtraction terms, (\ref{27}) and (\ref{27b}), decouple from the remaining contributions and give rise, by themselves, to a finite geometric conserved tensor. Using  the expressions in Appendix B for the different geometric quantities of a FLRW spacetime in terms of the scale factor, this conserved tensor turns out to be  
  \bea
\left< T_{\mu\nu}\right>_{Ad}^{(4)}=\frac{2}{2880\pi^2}\left[ -\frac{1}{2}\, ^{(1)}H_{\mu\nu}+\frac{11}{2}\,^{(3)}H_{\mu\nu}\right]  \ , \label{18}
\eea 
where 
\bea
^{(1)}H_{\mu\nu} & = & 2 R_{;\mu\nu}-2\Box R g_{\mu\nu}+2 RR_{\mu\nu}-\frac{1}{2}R^2 g_{\mu\nu} \ ,  \label{19}\\
^{(3)}H_{\mu\nu} & = &  R_{\mu}^{\rho}R_{\rho\nu}-\frac{2}{3} RR_{\mu\nu}-\frac{1}{2}R_{\rho\sigma}R^{\rho\sigma}g_{\mu\nu}+\frac{1}{4}R^2 g_{\mu\nu}  \ . \label{20}
\eea

 Therefore, we get the following expression for  the renormalized energy density and pressure: 
\bea
\left< T_{00}\right>_{ren}=\frac{1}{2\pi^2 a^3}\int_{0}^{\infty} dk k^2 && \left[  i \left( h_k^{I} \frac{\partial h_k^{I*}}{\partial t} + h_k^{II} \frac{\partial h_k^{II*}}{\partial t} - h_k^{I*} \frac{\partial h_k^{I}}{\partial t}  - h_k^{II*} \frac{\partial h_k^{II}}{\partial t} \right) \right. \nonumber \\ &&
\left. +2\omega + \frac{m^4 \dot{a}^2}{4 \omega^5 a^2} - \frac{m^2 \dot{a}^2}{4 \omega^3 a^2} \right] + \left< T_{00}\right>_{Ad}^{(4)} \ , \label{t00ren}
\eea
and 
\bea
\left< T_{ii}\right>_{ren}=\frac{-1}{2\pi^2 a}\int_{0}^{\infty}  & dk & k^2  \left[\frac{2k }{3a}[h_k^{I} h_k^{II*}+h_k^{I*} h_k^{II}]  -\frac{2}{3}\left[\omega-\frac{m^2}{\omega}\right]   \right.  \nonumber\\ 
 & - & \left.  \frac{m^2 \dot a^2}{12 \omega^3 a^2}-\frac{m^2 \ddot a}{6 \omega^3 a}+ \frac{m^4 \ddot a}{6 \omega^5 a}+\frac{m^4 \dot a^2}{2 \omega^5 a^2}-\frac{5 m^6 \dot a^2}{12 \omega^7 a^2}\right]  + \left< T_{ii}\right>_{Ad}^{(4)} \ , \label{tiiren} \eea
where the functions $(h_k^{I}, h_k^{II})$ above are exact solutions to Eq. (\ref{3b}) and provide the mode functions defining the quantum state.
Using (\ref{t00ren}) and (\ref{tiiren}) with (\ref{17f}), it is easy to see that, in the massless limit, the trace of the above tensor turns out to be 
\be \label{traceanomaly}\left< T^{\mu}_{\mu}\right>_{ren}= \left< T^{\mu}_{\mu}\right>_{Ad}^{(4)} =  \frac{2}{2880\pi^2}\left[ -\frac{11}{2}\left(R_{\mu\nu}R^{\mu\nu}-\frac{1}{3}R^2\right)+3\Box R\right] \ , \ee
in exact agreement with the conformal anomaly computed by other renormalization procedures and with the expression obtained in \cite{landete}.

Before seeing some examples of this formalism, we would like to discuss briefly the interpretation of the subtraction terms in terms of renormalization of constants in the gravitational action, and the potential ambiguities of the renormalization algorithm. As we have seen, the integrals of the zeroth and second order adiabatic subtractions in (\ref{Adsubtractions}) and (\ref{iiad}) do contain divergences. Following the procedure of \cite{Bunch80}, we can isolate them using dimensional regularization ($n$ is the spacetime dimension). We obtain
\bea -\frac{1}{2 \pi^2 a^3} \int_0^{\infty} dk k^2  \rho_k^{(0)}  &\rightarrow & -\frac{1}{2 \pi^2 a^3} \int_0^{\infty} dk k^{n-2}  \rho_k^{(0)} \approx\frac{m^4}{8 \pi^2} \frac{1}{n-4}+O(n-4) 
 \ ,  \label{000}  \\
 -\frac{1}{2 \pi^2 a^3} \int_0^{\infty} dk k^2  \rho_k^{(2)}  &\rightarrow & -\frac{1}{2 \pi^2 a^3} \int_0^{\infty} dk k^{n-2}  \rho_k^{(2)}\approx\frac{m^2}{8 \pi^2} \frac{1}{n-4}\frac{\dot a^2}{a^2} +O(n-4)
 \ ,  \label{ii0}  \eea 
and
\bea 
 -\frac{1}{2 \pi^2 a} \int_0^{\infty} dk k^2  p_k^{(0)}  &\rightarrow & -\frac{1}{2 \pi^2 a} \int_0^{\infty} dk k^{n-2}  p_k^{(0)} \approx -\frac{m^4 a^2}{8 \pi^2} \frac{1}{n-4}+O(n-4) 
 \ ,   \label{002}   \\
 -\frac{1}{2 \pi^2 a} \int_0^{\infty} dk k^2  p_k^{(2)}  &\rightarrow & -\frac{1}{2 \pi^2 a} \int_0^{\infty} dk k^{n-2}  p_k^{(2)} \approx -\frac{m^2 a^2}{24 \pi^2} \frac{1}{n-4} \left(2\frac{\ddot a}{a}+\frac{\dot a^2}{a^2}\right)+O(n-4)
  \ .  \label{ii2}  
\ \ \ \ \eea 

The  zeroth and second order adiabatic subtraction terms, (\ref{000}) and (\ref{ii0}) and also (\ref{002}) and (\ref{ii2}), are formally divergent, but can be suitable expressed  as  geometric tensors. Using the geometric identities of Appendix B one can easily find
\bea
\left< T_{\mu\nu}\right>_{Ad}^{(0)} & \approx &\frac{m^4}{8 \pi^2}\frac{1}{n-4} g_{\mu\nu}+O(n-4)  \ , \label{ren0}  \\
\left< T_{\mu\nu}\right>_{Ad}^{(2)} &\approx&-\frac{m^2}{24 \pi^2}\frac{1}{n-4}G_{\mu\nu}+O(n-4)  \label{ren2}  \ ,
\eea
with $G_{\mu\nu}=R_{\mu\nu}-\frac{1}{2}g_{\mu\nu}R$, the Einstein tensor. 

Recall now the Einstein's gravitational field equations
\bea
G_{\mu\nu}+\Lambda g_{\mu\nu}= -8\pi G  \left< T_{\mu\nu} \right> \ . \label{einstein}
\eea
Equation (\ref{ren0})  suggests the possibility of absorbing the UV divergence of the zeroth adiabatic order into the cosmological constant, $\Lambda$,  while expression (\ref{ren2}) offers the possibility of renormalizing the second adiabatic order divergence into the Newton's universal constant, $G$. This way, the adiabatic subtraction terms can be nicely interpreted in terms of renormalization of coupling constants, in parallel to the scalar fields case \cite{Bunch80}. 

On the other hand, as we have stressed before, in a general spacetime the fourth-order  subtraction terms give rise to proper UV divergencies \cite{Christensen78}. They turn out to be proportional to a linear combination of the two independent geometric tensors with the appropriate dimensions, namely $\ ^{(1)}H_{\mu\nu}$ and $\ ^{(2)}H_{\mu\nu} $. The four types of divergent subtraction terms, proportional to  $ m^4g_{\mu\nu}, m^2G_{\mu\nu}, \ ^{(1)}H_{\mu\nu}$ and $ \ ^{(2)}H_{\mu\nu}$,  generate intrinsic ambiguities in the curved space renormalization program for the stress-energy tensor \cite{birrell-davies, Waldbook}. The first two can be naturally associated to the renormalization of the cosmological constant and Newton's constant. In our FLRW spacetime $\ ^{(2)}H_{\mu\nu}$ is proportional to $\ ^{(1)}H_{\mu\nu}$, and hence we are left with only one relevant renormalization parameter ambiguity.  This translates to the fact that the general expression for the finite fourth order adiabatic contribution $\left< T_{\mu\nu}\right>_{Ad}^{(4)}$ contains actually an arbitrary coefficient $c_1$:
 \bea
\left< T_{\mu\nu}\right>_{Ad}^{(4)}=\frac{2}{2880\pi^2}\left[ c_1\, ^{(1)}H_{\mu\nu}+\frac{11}{2}\,^{(3)}H_{\mu\nu}\right]  \ . \label{18b}
\eea
Our adiabatic regularization method leads to $c_1=-1/2$. Other renormalization methods can only potentially differ from our results on the value of this coefficient. However, we remark that the ambiguity disappears for spacetime backgrounds for which the tensor $^{(1)}H_{\mu\nu}$ vanishes. This happens for physically relevant spacetimes, like de Sitter space or the radiation-dominated universe.


Note that if one considers the general expression (\ref{18b}), instead of (\ref{18}),
the numerical coefficient of $\Box R$ in (\ref{traceanomaly}) is actually proportional to $c_1$.


 \subsection{Stress-energy conservation}

The above $\left< T_{\mu \nu}\right>_{ren}$ is, as expected, a conserved tensor $\nabla^{\mu}\left< T_{\mu\nu}\right>_{ren}=0$.  The conservation equations in a FLRW spacetime can be spelled out as 
\bea
\langle T^{0\nu}_{\hspace{0.3 cm};\nu} \rangle & = & \langle \dot T_{00} \rangle +3\frac{\dot a}{a} \langle T_{00} \rangle +\frac{3}{a^2}\frac{\dot a}{a} \langle T_{ii} \rangle = 0 \ ,  \\
\langle T^{i \nu}_{\hspace{0.3 cm};\nu} \rangle & = & 0, \hspace{2cm} i=1,2,3
\eea
and they can be checked by  direct computation. This is a consequence of the fact that, for each adiabatic order of the formal subtraction tensors $\langle T_{\mu\nu}\rangle_{Ad}^{(n)}$,
where 
\be \langle T_{00}\rangle_{Ad}^{(n)}= -\frac{1}{2\pi^2 a^3}\int_0^{\infty} dk k^2 \rho_k^{(n)} \ , \ee
\be \langle T_{ii}\rangle_{Ad}^{(n)}= -\frac{1}{2\pi^2 a}\int_0^{\infty} dk k^2 p_k^{(n)} \ , \ee
we have the independent  conservation laws $\nabla^{\mu}\left< T_{\mu\nu}\right>_{Ad}^{(n)}=0$, for $n=0,2,4$.

\section{Examples}

In this section, we work out the renormalized stress-energy tensor for two different spacetimes: de Sitter spacetime and a radiation-dominated universe. For a given scale factor $a(t)$, we need to solve (\ref{3b}). As it is a system of two coupled first-order differential equations, we have 2 degrees of freedom which have to be fixed somehow. More specifically, given a particular solution $(h_k^I, h_k^{II})$ normalized as in (\ref{4}),  we can construct the general solution by a Bogolubov-type rotation: 
\bea
h_k^{I} & \to & E_k h_k^{I} + F_k h_k^{II*} \ , \nonumber \\
h_k^{II} & \to & E_k h_k^{II} - F_k h_k^{I*}  \label{bogolubov-sol} \ , 
\eea 
where $E_k$ and $F_k$ are two arbitrary constants. On the other hand, we should also ensure the normalization condition (\ref{4}), which implies the following constraint: \be |E_k|^2 + |F_k|^2=1 \  . \ee 
Note  that the renormalization ambiguity associated to the $^{(1)}H_{\mu\nu}$ tensor disappears for de Sitter spacetime and the radiation-dominated universe.

\subsection{Renormalized stress-energy tensor in de Sitter spacetime}
For de Sitter spacetime $a(t) = e^{H t}$ with $H$ a constant, the general solution to the field equations (\ref{3c}) and (\ref{3d}) can be conveniently expressed, using the transformation in (\ref{bogolubov-sol}), as the following linear combination:
\bea \label{dsgs}h_k^I (t) &=& E_k \left (\frac{i}{2} \sqrt{\pi z} e^{\frac{\pi \mu}{2}} H^{(1)}_{\frac{1}{2} - i \mu} (z)\right ) + F_k \left (\frac{1}{2} \sqrt{\pi z} e^{\frac{\pi \mu}{2}} H^{(1)}_{-\frac{1}{2} - i \mu} (z)\right )^* \ ,  \\
h_k^{II} (t) &=& E_k \left ( \frac{1}{2} \sqrt{\pi z} e^{\frac{\pi \mu}{2}} H^{(1)}_{-\frac{1}{2} - i \mu} (z)\right ) - F_k \left (\frac{i}{2} \sqrt{\pi z} e^{\frac{\pi \mu}{2}} H^{(1)}_{\frac{1}{2} - i \mu} (z)\right )^* \ , \label{dsgs2}\eea
where $z \equiv k H^{-1} e^{-H t}$, $\mu \equiv m/H$, $H^{(1)} (z)$ are Hankel functions of the first kind and $E_k$ and $F_k$ are constants that need to be fixed with appropriate initial conditions.
A crucial physical requirement is that, as $k \to \infty$,  the physical solutions should have the adiabatic asymptotic form
\be \label{adiabatic-k}h_k^I \sim \sqrt{\frac{\omega + m}{2\omega}} e^{-i \int^{t'} \omega (t') dt'} \ , \hspace{2cm} h_k^{II} \sim \sqrt{\frac{\omega - m}{2\omega}} e^{-i \int^{t'} \omega (t') dt'}   \ . \ee
This way, one recovers in this limit the Minkowskian solutions. This leads to
\be E_k \sim 1\ , \hspace{2cm}  F_k \sim 0  \ , \label{ekfkds}\ee
as $ k \to \infty$. 
The above condition can be naturally achieved by the simple solution $E_k=1$ and $F_k=0$. This determines a vacuum for spin-1/2 fields analogous to the Bunch-Davies vacuum \cite{BD} for scalar fields. It is also the natural extension of the conformal vacuum for massless fields. It can be uniquely characterized by invoking de Sitter invariance \cite{parker-toms}.
  
By changing the integration variable from $k$ to $z$, we obtain that the energy density (\ref{ren-ten}) is
\be \langle T_{00} \rangle_{ren} = \frac{H^3}{2 \pi^2} \int_0^{\infty} dz z^2 (\rho_k - \rho_k^{(0)} - \rho_k^{(2)} - \rho_k^{(4)} ) \label{ren-den-ds} \ , \ee
where from (\ref{dsgs}), (\ref{dsgs2}) and (\ref{ekfkds}), the bare contribution (\ref{13}) is
\be \rho_k (z) =  \frac{\pi H e^{\pi \mu} z^2}{2}   \Big(\frac{\mu}{z}\left[  H_{\nu-1}^{(1)}H_{-\nu}^{(2)}-H_{\nu}^{(1)}H_{1-\nu}^{(2)} \right] +i\left[  H_{\nu-1}^{(1)}H_{1-\nu}^{(2)}-H_{\nu}^{(1)}H_{-\nu}^{(2)} \right] \Big) \ , \ee
with $\nu \equiv (1/2) - i \mu$ and $H_{\nu}^{(1,2)} \equiv H_{\nu}^{(1,2)} (z)$. On the other hand, the subtraction terms, from (\ref{rho0}), (\ref{rho2}) and (\ref{17}), take the form
\be \rho_k^{(0)} = - 2 H \sqrt{z^2 + \mu^2} \ , \ee
\be \rho_k^{(2)} = H \left( \frac{\mu^2}{4 (z^2 + \mu^2)^{3/2}} - \frac{\mu^4}{4 (z^2 + \mu^2)^{5/2}} \right) \ ,\ee
\be \rho_k^{(4)} = H \left( \frac{105 \mu^8}{64 (z^2 + \mu^2)^{11/2}} - \frac{119 \mu^6}{32 (z^2 + \mu^2)^{9/2}} + \frac{165 \mu^4}{64 (z^2 + \mu^2)^{7/2}} - \frac{\mu^2}{2 (z^2 + \mu^2)^{5/2}} \right) \ .\ee
Similar expressions can be obtained for the renormalized pressure (\ref{17b}), 
\be \langle T_{ii} \rangle_{ren} = \frac{a^2 H^3}{2 \pi^2} \int_0^{\infty} dz z^2 (p_k - p_k^{(0)} - p_k^{(2)} - p_k^{(4)} ) \label{ren-pre-ds} \ , \ee
where, from (\ref{13d}),
\bea
p_k(z)=i \frac{\pi H e^{\pi \mu}z^2}{6}\left[ H_{\nu-1}^{(1)}H_{1-\nu}^{(2)}-H_{\nu}^{(1)}H_{-\nu}^{(2)} \right] \ .
\eea
From these results, one can obtain $\langle T_{00} \rangle_{ren}$ and $\langle T_{ii} \rangle_{ren}$ numerically with very high accuracy.   We reproduce exactly the  analytical expression  already obtained in \cite{landete} from the trace anomaly and the symmetries of de Sitter spacetime
\be \langle T_{\mu \nu} \rangle _r = \frac{1}{960 \pi^2} g_{\mu \nu} \left[11 H^4 + 130 H^2 m^2 + 120 m^2 (H^2 + m^2) \left( \log \left(\frac{m}{H} \right) - \mathfrak{Re} \left[ \psi \left( - 1 + i \frac{m}{H} \right) \right] \right) \right]  \label{desitterten} \ , \ee
where $\psi (z)$ is the digamma function.

\subsection{Radiation dominated universe}

In this section, we apply our general results for a radiation-dominated universe. This  is also a nice example to show  how the  general procedure works.  In this case, the  two independent solutions of the differential equation for the field modes with $a(t)= a_0 \sqrt{t}$ are given in terms of the Whittaker functions $\frac{1}{a(t)} W_{\kappa,\mu}(z)$ and $\frac{1}{a(t)}W_{-\kappa,\mu}(-z)$ (see \cite{Barut-Duru}), where 
\bea 
\kappa= \frac{1}{4}-ix^2,  \hspace{0.5cm} \mu=\frac{1}{4}, \hspace{0.5cm} z=i 2mt  \ , \label{32}
\eea
with $  x^2\equiv k^2 / (a_0^2 2m)$.
We choose a set of two linear independent solutions for the field modes of the form
\bea
h_k^{I}= E_k \left ( N \frac{W_{\kappa,\mu}(z)}{\sqrt{a(t)}} \right )    
+ F_k \left ( N \frac{k}{2m a(t)^{3/2} }\left[W_{\kappa,\mu}(z)+\left(\kappa-\frac{3}{4}\right)W_{\kappa-1,\mu}(z) \right]   \right )^* \ , \label{34}
\eea
where the constant $N=\frac{a_0^{1/2}}{(2m)^{1/4}}e^{-\frac{\pi}{2}x^2}$ and the condition $|E_k|^2 + |F_k|^2=1$ are fixed from the normalization condition (\ref{4}).
The adiabatic condition (\ref{adiabatic-k}) for $k \to \infty$ also requires that $E_k \sim 1$ and $F_k \sim 0$. Moreover, 
a detailed analysis (see Appendix C) of the asymptotic properties of the stress energy tensor components using the Whittaker functions \cite{lopez} allows us to characterize the condition for the renormalizability of the vacuum expectation values of the stress tensor as
\be \label{renor} |E_k|^2 - |F_k|^2 = 1 + O(k^{-5}) \ . \ee
As one can see from Appendix C, this particular combination of $E_k$ and $F_k$ is the crucial one in the analysis of the renormalized energy density (\ref{37}) and pressure (\ref{37b}) for 1/2 spin fields.

In contrast with the previous example of de Sitter spacetime, the absence of  extra symmetries (in addition to the standard homogeneity and isotropy of a FLRW spacetime) for the radiation-dominated background  does not allow us to fix a natural  preferred vacuum state. 
However, the early and late-time behaviors ($t << m^{-1}$ and $t>> m^{-1}$, respectively) of the renormalized stress-energy tensor can be obtained generically, and agree with the forms assumed by classical cosmology. As detailed in Appendix C, we have that,
as time evolves and reaches the regime $ t >> m^{-1}$, the renormalized energy density takes the form of cold matter 
\be \left<T_{00} \right>_{ren} \sim \frac{\rho_{0m}}{a^3} \ , \ee where
\bea
\rho_{0m} = \frac{m}{ \pi^2}\int_0^{\infty} dk k^2 \left[1-(|E_k|^2-|F_k|^2) \right] \geq 0 \ . 
\eea
Notice that, $2\geq 1-(|E_k|^2-|F_k|^2)=2 |F_k|^2\geq 0$, and together with the renormalizability condition (\ref{renor}), we see that the energy density $\rho_{0m}$ is finite and definite positive. The specific value of $\rho_{0m}$ depends on the form of the quantum state for our spin-$1/2$ field, i.e. of the choice of $E_k$ and $F_k$. Since at late times $ t >> m^{-1}$ the relation (\ref{34}) transforms into
 \be h^{I}_{{k}}(t) \sim E_k \sqrt{\frac{\omega + m}{2\omega}} e^{-i \int^{t'} \omega (t') dt'}  +
F_k \sqrt{\frac{\omega - m}{2\omega}} e^{i \int^{t'} \omega (t') dt'}\ ,  \ee 
the coefficients $F_k$ are actually the fermionic $\beta$-type (Bogolubov) coefficients \cite{parker66, parker68}. Therefore, we actually get  
 $\rho_{0m} \sim m \langle n(t) \rangle$, where $\langle n(t) \rangle$ is the number density of the  created  particles.
  Moreover,  we find  
 \be \frac{\left< T_{ii} \right>_{ren}}{a^2} \sim 0 \ , \ee and hence the pressure obeys the cold matter equation of state.  Note that the potential renormalization ambiguity, proportional to $\ ^{(1)}H_{\mu\nu}$, is here identically zero. In more general cosmological spacetimes these fourth-order adiabatic terms rapidly vanish at late-times.

On the other hand, for sufficiently early times in the evolution, $t << m^{-1}$, we have  (see Appendix C)
\be \left<T_{00} \right>_{ren} \sim \frac{\rho_{0r}}{a^4} \ , \ee 
with
\be \rho_{0r} =  \frac{1}{ \pi^2} \int_0^{\infty} dk k^3\left[1-(|E_k|^2-|F_k|^2)\right] \geq 0\ , 
\ee
and additionally
\be \frac{\left<T_{ii} \right>_{ren}}{a^2} \sim \frac{1}{3} \left< T_{00} \right>_{ren} \label{Tii2/3} \ . \ee
Note again that $\rho_{0r}$ is finite and definite positive. The specific value of $\rho_{0r}$ depends on the specific form of the quantum state throughout the complex functions $E_k$ and $F_k$. From (\ref{Tii2/3}), we see that $p \sim \rho/3$, in agreement with the assumptions 
of classical cosmology for the radiation. 

The analysis and phenomenology of the renormalized stress-energy tensor obtained from specific choices of the vacuum state is beyond the scope of the present paper. Note again that any choice for the quantum state has not  a preferred status, in contrast with  the  Bunch-Davies type vacuum of the previous example, due to the absence of the additional symmetries  endowed by de Sitter spacetime.

\section{Conclusions}

In this paper, we have extended the adiabatic regularization method to the renormalization of the spin-1/2 stress-energy tensor. The main result of this paper is Eqs. (\ref{t00ren}) and (\ref{tiiren}), which provide expressions that are simple and numerically easy to compute, once the quantum state  is given. We have illustrated our approach by briefly analyzing de Sitter space, with an assumed Bunch-Davies type vacuum state. The renormalized energy and pressure densities coincide with those predicted by symmetry arguments \cite{landete}. We have also analyzed the renormalized stress-energy tensor in a purely radiation-dominated universe. In the latter case the early and late-time behavior of the renormalized stress-energy tensor can be worked out explicitly, irrespective of the specific form of the quantum state, and agree with those assumed by classical cosmology for radiation and cold matter, respectively.
\newline

\section*{Acknowledgments}
  
This work is supported  by the  Research Projects of the Spanish MINECO   FIS2011-29813-C02-02,  FPA2012-39684-C03-02,  the Consolider Program No. CPANPHY-1205388, and the Centro de Excelencia Severo Ochoa Program No. SEV-2012-0249. F. T. is supported by the Severo Ochoa Ph.D. fellowship SVP-2013-067697.  A. D. is supported by the Generalitat Valenciana ValI+D Ph.D. fellowship ACIF/2014/070. J. N-S. would like to thank I. Agullo, G. Olmo and L. Parker for very useful discussions. 


\appendix

\section{Fermionic adiabatic expansion}

We give here the third and  fourth  order contributions to $F(t)$ and $\omega (t)$ in (\ref{adifexp}) [we have $G^{(n)} (m) = F^{(n)} (-m)$]:

{\small 
\begin{eqnarray}
\omega^{(3)}&=&0 \label{omega3}\ ,  \\
F^{\left(3\right)} & = & i \left( \frac{65 m^5 \dot{a}^3}{64 a^3 \omega^8} - \frac{97 m^3 \dot{a}^3}{128 a^3 \omega^6} + \frac{m \dot{a}^3}{16 a^3 \omega^4} - \frac{19 m^3 \dot{a} \ddot{a}}{32 a^2 \omega^6} + \frac{m \dot{a} \ddot{a}}{4 a^2 \omega^4} + \frac{m \dddot{a}}{16 a \omega^4} \right)\: \label{eq:singad3}  \ ,
\end{eqnarray}
}{\small \par}
and 
{\small \bea \omega^{(4)} = &-& \frac{1105 m^8 \dot{a}^4}{128 a^4 \omega^{11}} + \frac{337 m^6 \dot{a}^4}{32 a^4 \omega^9} - \frac{377 m^4 \dot{a}^4}{128 a^4 \omega^7} + \frac{ 3 m^2 \dot{a}^4}{32 a^4 \omega^5} + \frac{221 m^6 \dot{a}^2 \ddot{a}}{32 a^3 \omega^9} - \frac{389 m^4 \dot{a}^2 \ddot{a}}{64 a^3 \omega^7} + \frac{13 m^2 \dot{a}^2 \ddot{a}}{16 a^3 \omega^5} - \frac{19 m^4 \ddot{a}^2}{32 a^2 \omega^7} \nonumber \\
&+& \frac{m^2 \ddot{a}^2}{4 a^2 \omega^5} - \frac{7 m^4 \dot{a} \dddot{a}}{8 a^2 \omega^7} + \frac{15 m^2 \dot{a} \dddot{a}}{32 a^2 \omega^5} + \frac{m^2 \ddddot{a}}{16 a \omega^5} \ , \eea 
\bea 
F^{(4)} &=& + \frac{2285 m^8 \dot{a}^4}{512 a^4 \omega^{12}} - \frac{565 m^7 \dot{a}^4}{128 a^4 \omega^{11}} - \frac{1263 m^6 \dot{a}^4}{256 a^4 \omega^{10}} + \frac{2611 m^5 \dot{a}^4}{512 a^4 \omega^9} + \frac{2371 m^4 \dot{a}^4}{2048 a^4 \omega^{8}} - \frac{333 m^3 \dot{a}^4}{256 a^4 \omega^7} - \frac{3 m^2 \dot{a}^4}{128 a^4 \omega^6} + \frac{m \dot{a}^4}{32 a^4 \omega^5} \nonumber \\
&& - \frac{457 m^6 \dot{a}^2 \ddot{a}}{128 a^3 \omega^{10}} + \frac{113 m^5 \dot{a}^2 \ddot{a}}{32 a^3 \omega^9} + \frac{725 m^4 \dot{a}^2 \ddot{a}}{256 a^3 \omega^8} - \frac{749 m^3 \dot{a}^2 \ddot{a}}{256 a^3 \omega^7} - \frac{19 m^2 \dot{a}^2 \ddot{a}}{64 a^3 \omega^6} + \frac{11 m \dot{a}^2 \ddot{a}}{32 a^3 \omega^5} + \frac{41 m^4 \ddot{a}^2}{128 a^2 \omega^8} - \frac{5 m^3 \ddot{a}^2}{16 a^2 \omega^7} \nonumber \\
&& - \frac{17 m^2 \ddot{a}^2}{128 a^2 \omega^6} + \frac{m \ddot{a}^2}{8 a^2 \omega^5} + \frac{7 m^4 \dot{a} \dddot{a}}{16 a^2 \omega^8} - \frac{7 m^3 \dot{a} \dddot{a}}{16 a^2 \omega^7} - \frac{13 m^2 \dot{a} \dddot{a}}{64 a^2 \omega^6} + \frac{7 m \dot{a} \dddot{a}}{32 a^2 \omega^5} - \frac{m^2 \ddddot{a}}{32 a \omega^6} + \frac{m \ddddot{a}}{32 a \omega^5} \ . \label{effe4}
\eea }



\section{Useful formulas for a FLRW spacetime}

In  checking that the fourth order adiabatic subtraction terms (\ref{27}) and (\ref{27b})  give the covariant result (\ref{18}) we used the following results:
\bea
R_{00} & = & 3\frac{\ddot a}{a} \ ,  \hspace{2cm} R_{ij}=-a^2\left[2 \frac{\dot a^2}{a^2}+\frac{\ddot a}{a}  \right]\delta_{ij}  \label{21}\ , \\
R^2 & = & 36\left[\frac{\dot a^4}{a^4}+2\frac{\dot a^2}{a^2}\frac{\ddot a}{a}+\frac{\ddot a^2}{a^2} \right]  \ , \label{22}\\
\Box R & = & 6\left[ \frac{\ddot a^2}{a^2}+\frac{\ddddot a}{a}-5\frac{\dot a^2}{a^2}\frac{\ddot a}{a} +3\frac{\dot a}{a}\frac{\dddot a}{a}\right]\ ,   \label{23}\\
R_{\mu\nu}R^{\mu\nu} & = & 12\left[\frac{\dot a^4}{a^4}+\frac{\ddot a^2}{a^2}+\frac{\dot a^2}{a^2}\frac{\ddot a}{a} \right]  \ , \label{24}\\
R_{; 00} & = &  6\left[\frac{\ddddot a}{a}+\frac{\ddot a^2}{a^2}-8\frac{\dot a^2}{a^2}\frac{\ddot a}{a}+6\frac{\dot a^4}{a^4} \right] \ ,  \label{25}\\
R_{; ij} & = & -6 a^2\left[\frac{\dddot a}{a}\frac{\dot a}{a}+\frac{\ddot a}{a}\frac{\dot a^2}{a^2}-2\frac{\dot a^4}{a^4}  \right]\delta_{ij}  \ .  \label{26}
\eea

\section{Asymptotic analysis of the energy density and pressure for a spin 1/2 field in a radiation dominated universe}

In this Appendix we shall study the asymptotic properties of the stress tensor components in a radiation-dominated universe. The large momentum behavior of this tensor will give us a necessary and sufficient condition for its renormalizability, while the late/early-time behavior will reproduce the classical results of physics of fluids for a matter/radiation-dominated universe.

Recall the general solution (\ref{34}) for the modes in a radiation-dominated universe, and define for simplicity  the following quantities:
\bea
g_k^{I}(t) & \equiv & N \frac{W_{\kappa,\mu}(z)}{\sqrt{a(t)}} \ ,  \\
g_k^{II}(t) & \equiv &  N \frac{k}{2m a(t)^{3/2} }\left[W_{\kappa,\mu}(z)+\left(\kappa-\frac{3}{4}\right)W_{\kappa-1,\mu}(z) \right]   \ .
\eea
Expressions (\ref{13}), (\ref{13d}), and (\ref{17f}), can be rewritten in terms of these independent solutions by doing a a Bogolubov-type rotation $(h_k^{I} \to E_k g_k^{I} + F_k g_k^{II*},\ h_k^{II} \to E_k g_k^{II} - F_k g_k^{I*} )$
\bea
\rho_k  & = & \rho_k^{D}   \left[|E_k|^2-|F_k|^2 \right]  -E_k F_k^*\rho_k^{ND}-E_k^* F_k\rho_k^{ND*}  \ ,  \label{37}\\
p_k & = & p_k^{D} \left[|E_k|^2-|F_k|^2 \right] - E_k F_k^*p_k^{ND}  -E_k^*F_k p_k^{ND*}  \ ,  \label{37b}\\
\rho_k^D & = & 3p_k^D-2m\left[|g_k^I|^2-|g_k^{II}|^2 \right] \ ,   \label{37c}
\eea
where
\bea
\rho_k^{D} & = &i\left[g_k^I \frac{\partial g_k^{I*}}{\partial t}+g_k^{II} \frac{\partial g_k^{II*}}{\partial t}-g_k^{I*} \frac{\partial g_k^{I}}{\partial t}-g_k^{II*} \frac{\partial g_k^{II}}{\partial t}\right] \ ,  \label{37d}\\
\rho_k^{ND} & = & -2i\left[g_k^I \frac{\partial g_k^{II}}{\partial t}-g_k^{II} \frac{\partial g_k^{I}}{\partial t}\right]  \ ,  \label{37e}\\
p_k^{D} & = & -\frac{2k}{3a}\left[g_k^{I} g_k^{II*}+g_k^{I*} g_k^{II} \right] \ ,   \label{37f}\\
p_k^{ND} & = & -\frac{2k}{3a}\left[(g_k^{I})^2-(g_k^{II})^2\right] \ . \label{37g}
\eea
The energy density and the pressure functions, expressed  this way, show explicitly the dependence on the vacuum state.

Using the result (\ref{37d}), and derivative and functional properties of the Whittaker functions \cite{handbook}, one can find for the energy density,
\bea
\rho_k^D& =&  -2m  - \frac{4m x^2e^{-\pi x^2}}{|z|^{3/2} }    \label{38} \\
 &\times &\left[ |W_{\kappa,\frac{1}{4}}(z)|^2-\left(\frac{1}{4}+x^4 \right)|W_{\kappa-1,\frac{1}{4}}(z)|^2  \right]\nonumber \ . 
\eea 
For large values of the momenta, $k^2/a^2>>m^2$, the Whittaker function can be very well approximated by \cite{lopez}
\begin{widetext}
\bea
W_{\kappa,1/4}(z) &= & \frac{ \sqrt{\pi} z^{1/4}}{\Gamma\left(\frac{3}{4}-\kappa\right)} \left\{ \cos(2\sqrt{\kappa z})\left[1-\frac{M_2(z)}{\kappa}+\frac{M_4(z)}{\kappa^2}- \left(\frac{M_1(z)}{\kappa}-\frac{M_3(z)}{\kappa^2}\right)\frac{\Gamma(3/4-\kappa)}{\Gamma(1/4-\kappa)} \right] \right.   \label{38bb} \\
 & - & \left.  \frac{\sin(2\sqrt{\kappa z})}{\sqrt{\kappa}}\left[M_1(z)-\frac{M_3(z)}{\kappa}+\frac{M_5(z)}{\kappa^2}+\left(1-\frac{M_2(z)}{\kappa}+\frac{M_4(z)}{\kappa^2}\right)\frac{\Gamma(3/4-\kappa)}{\Gamma(1/4-\kappa)} \right] \right\} +O(|\kappa|^{-3})\nonumber \ , 
\eea 
\end{widetext}
where $M_n(z)$ are a set of polynomials that satisfy 
\bea
M_1(z) & = & -\frac{z^{3/2}}{12}\ ,   \label{38bc}\\
M_2(z) & = & -\frac{z}{16}\left(1-\frac{z^2}{18}\right) \ ,  \label{38bd}\\
M_3(z) & = & -\frac{z^{1/2}}{32}+\frac{z^{5/2}}{120}-\frac{z^{9/2}}{10368}  \ ,  \label{38be}\\
M_4(z) & = & -\frac{1}{128}+\frac{19z^{2}}{1536}-\frac{11z^{4}}{23040}+\frac{z^{6}}{497664}  \ , \label{38bf}\\
M_5(z) & = & \frac{z^{3/2}(2721600 - 291924 z^2 + 3528 z^4 - 7 z^6)}{209018880}  \label{38bg} \ .
\eea
After a long calculation one can find that 
\bea
|W_{\kappa,\frac{1}{4}}(z)|^2 & - & \left(\frac{1}{4}+x^4 \right)|W_{\kappa-1,\frac{1}{4}}(z)|^2   \label{38bh}\\
 & = &  \left(\frac{|z|}{x}-\frac{|z|^{3/2}}{2x^2}+\frac{|z|^{2}}{8x^3}-\frac{|z|(1+|z|^2)}{128 x^5}+O(x^{-7}) \right)e^{\pi x^2} \ ,  \nonumber
 \eea
so that, taking the change $x=\sqrt{\frac{t}{2m}(\omega^2-m^2)}$, one gets
\bea
\rho_k^D= \left[-2\omega+\frac{m^2}{16t^2\omega^3}+O(\omega^{-5})\right]  \ . \label{38bi}
\eea
From this expression it is easy to see that we recover those terms of zeroth and second adiabatic order found in Eqs. (\ref{rho0})-(\ref{rho2}) for a radiation-dominated universe. These contributions give the divergences of the stress-energy tensor. Additionally, the no-diagonal terms are shown to be
\bea
\rho_k^{ND} & = & O(\omega^{-1})  \ . \label{NDrho}
\eea
On the other hand, taking (\ref{37f}) [or more easily (\ref{37c})] we can find for the pressure
\bea
p_k^D & = & -\frac{4m}{3}  -\frac{4m x^2e^{-\pi x^2}}{3|z|^{3/2} }   \nonumber\\
 &\times &\left[\left(1-\frac{|z|}{x^2}\right) |W_{\kappa,\frac{1}{4}}(z)|^2-\left(\frac{1}{4}+x^4 \right)|W_{\kappa-1,\frac{1}{4}}(z)|^2  \right]   \nonumber\\
 & = & -\left[\frac{2}{3}\omega-\frac{2m^2}{3\omega}-\frac{m^2}{48t^2\omega^{3}}+O(\omega^{-7})\right]    \label{38bj} \ , 
\eea
which also agrees with the divergences found in (\ref{17c}) and (\ref{17d}). Additionally,
\bea
p_k^{ND} & = & O(\omega^{-1})  \ . \label{NDp}
\eea

The choice of the parameters $E_k$ and $F_k$ (the choice of the vacuum state) is determined by imposing some initial condition at a given instant of time, $t_0$. This choice must be in such a way that leaves the stress energy tensor without divergences. According to (\ref{38bi}), (\ref{38bj}) and (\ref{ren-ten}), (\ref{17b}) respectively, the stress energy tensor renormalizability imposes a natural constraint on the vacuum  state [recall (\ref{37}) and (\ref{37b})],
\bea
|E_k|^2-|F_k|^2= 1+ O(\omega^{-5})  \ .   \label{38bja}
\eea
This means that $E_k=1+O(\omega^{-5}) $ and $F_k=O(\omega^{-5/2})$, which makes $E_k F_k^*=O(\omega^{-5/2})$, and it is enough for the no-diagonal terms, (\ref{NDrho}) and (\ref{NDp}),  to not to give new divergences.

Let us focus now on the stress-energy tensor for late times in the expansion of the universe. 
Taking $t >> m^{-1}$, equation (\ref{38}) behaves as
\bea
\rho_k^D=-2m-\frac{4m x^2}{|z|}+\frac{4 m x^4}{|z|^2}+\dots \ , 
\eea
while
\bea
\rho_k^{(0)}+\rho_k^{(2)}+\rho_k^{(4)}= \rho_k^D+O(|z|^{-7}) \ , 
\eea
so we may state, recalling (\ref{ren-ten}),
\bea
\left<T_{00} \right>_{ren}(t>> m^{-1})=\frac{1}{2 \pi^2 a^3}\left[\int_0^{\infty} dk k^2 2m\left[1-(|E_k|^2-|F_k|^2) \right]+O(|z|^{-1})\right] \ .  \label{latetimesdensity}
\eea
 Similarly, one can study the late-times behavior of the pressure (\ref{38bj}), and find
\bea
p_k^D=-\frac{8m x^2}{3 |z|}+\frac{16 m x^4}{3 |z|^2}-\frac{2m x^2(-1+8x^4)}{|z|^3}+\dots \ , 
\eea
while the corresponding adiabatic subtractions are
\bea
p_k^{(0)}+p_k^{(2)}+p_k^{(4)}= p_k^D+O(|z|^{-7}) \ .
\eea
Again, following (\ref{17b}) we find
\bea
\left<T_{ii} \right>_{ren}(t >> m^{-1})=\frac{1}{2 \pi^2 a}\left[ \int_0^{\infty} dk k^2 \frac{8mx^2}{3|z|}\left[1-(|E_k|^2-|F_k|^2) \right]+O(|z|^{-2})\right] \ . \label{prematter}
\eea
 This time, the dominant contribution to the total pressure, $p\equiv \left<T_{ii} \right>_{ren}/a^2$, decays with time. Basically, Eqs. (\ref{latetimesdensity}) and (\ref{prematter}) tell us that in a radiation-dominated expansion of the universe, a spin-1/2 field tends to behave as a source of  cold matter in cosmology. This may be useful to analyze in detail the phase transition from radiation to matter dominated universes, in the standard cosmology.
 
On the other hand, at early times $t<<m^{-1}$, (\ref{38}) reads [we analyze only the large momentum behavior since it is in this case where any problem with divergences might arise]
\bea
\rho_k^D & = &  i\frac{ 4m \pi e^{-\pi x^2}}{\sqrt{|z|}} \left[\frac{(-1)^{1/4}}{\Gamma(i x^2)\Gamma(1/2-ix^2)}+\frac{(-1)^{3/4}}{\Gamma(-i x^2)\Gamma(1/2+ix^2)} \right]+O(|z|^{0})\\
 & = & \frac{1}{\sqrt{|z|}}\left[-4m x+\frac{m}{32x^3}+\frac{21m}{8192x^7}+O(x^{-11})\right]+O(|z|^{0}) \label{final1}\ ,
\eea
and just as in the late-time case we obtain
\bea
\rho_k^{(0)}+\rho_k^{(2)}+\rho_k^{(4)}= \frac{1}{\sqrt{|z|}}\left[-4m x+\frac{m}{32x^3}+\frac{21m}{8192x^7}\right]+O(|z|^{1/2}) \label{final2}\ ,
\eea
so at early times,
\bea
\left<T_{00} \right>_{ren}(t << m^{-1})\approx \frac{1}{2 \pi^2 a^3}\left[\int_0^{\infty} dk k^2 \frac{4mx}{\sqrt{|z|}}\left[1-(|E_k|^2-|F_k|^2) \right]+O(|z|^{0})\right] \ .
\eea
Finally, if one tries to do the same calculation to the pressure (\ref{38bj}), one finds just the same results for (\ref{final1}) and (\ref{final2}) but with a factor $1/3$, so recovering this way the equation of state for classical radiation in cosmology.

\end{document}